\documentclass[11pt,a4paper]{article}

\usepackage{epsfig}
\usepackage{latexsym}
\usepackage{amsfonts}
\usepackage{amsmath}
\usepackage{amsthm}
\usepackage{amssymb}
\usepackage{amsbsy}
\usepackage{multirow}
\usepackage{slashed}
\usepackage{color}
\usepackage{mathrsfs}
\usepackage{subfigure}
\usepackage[font={footnotesize,sl},bf]{caption}
\usepackage{wrapfig}
\usepackage{jheppub}

\newcommand\be{\begin{equation}}
\newcommand\bea{\begin{eqnarray}}
\newcommand\ee{\end{equation}}
\newcommand\eea{\end{eqnarray}}

\newcommand{\bdm}{\begin{displaymath}}
\newcommand{\edm}{\end{displaymath}}

\newcommand{\ket}[1]{|#1 \rangle}
 
\title{Is Alice burning or fuzzing?}

\preprint{IPhT-t12/063\\}

\author[1]{Borun D.\ Chowdhury}
\author[2]{, Andrea Puhm}

\affiliation[1]
{Institute for Theoretical Physics, University of Amsterdam,\\
Science Park 904, Postbus 94485, 1090 GL Amsterdam, The Netherlands}
\affiliation[2]
{Institut de Physique Th\'eorique, CEA Saclay, 91191 Gif sur Yvette, France}

\abstract{Recently, Almheiri, Marolf, Polchinski and Sully (AMPS) have suggested a
Gedankenexperiment to test black hole complementarity. They claim that the postulates
of black hole complementarity are mutually inconsistent and choose to give up the `absence
of drama' for an infalling observer. According to them the black hole is shielded by a
firewall no later than Page time. This has generated some controversy. We find that
an interesting picture emerges when we take into account objections from the advocates
of fuzzballs. We reformulate AMPS' Gedankenexperiment in the decoherence picture of
quantum mechanics and find that low energy wave packets interact with the radiation
quanta rather violently while high energy wave packets do not. This is consistent with
Mathur's recent proposal of \emph{fuzzball complementarity} for high energy quanta
falling into fuzzballs.
}

\begin{document}
\maketitle

\section{Introduction} \label{Intro}

Recently Almheiri, Marolf, Polchinski and Sully (AMPS) have argued  that if a black hole formed by  collapse of a pure state is to evaporate away to a pure state, then an observer falling into the black hole at sufficiently late times will encounter high energy quanta and burn up at a {\it firewall} \cite{Almheiri:2012rt}.

To many people this is a surprising result because of the lore that the state at the horizon\footnote{We do not attempt go give a precise definition here; we think the concept of a `state at the horizon' is an ill-defined concept  in a full theory of quantum gravity where the horizon cannot exist according to the fuzzball proposal as it leads to information loss.} of black holes formed by collapse of a pure state is the vacuum state for the  infalling observer. 
However, previous papers of Mathur \cite{Mathur:2009hf,Mathur:2011uj} and Avery \cite{Avery:2011nb} have shown that small corrections (those which vanish for very massive black holes) to such a state cannot ensure purity of the final  state.
They used these results to argue for the fuzzball proposal.
While, the recent Gedankenexperiment by AMPS has incorporated this result and provided a testing ground for black hole complementarity, they conclude that these degrees of freedom at the horizon are a firewall that leads to a `drama' for the infalling observer.

There have been several responses to AMPS' firewall result. In \cite{Mathur:2012jk} Mathur and Turton claim that a macroscopic detector cannot detect the difference between the thermal and a typical state before crossing the horizon. 
In \cite{Bousso:2012as,Harlow:2012me}\footnote{While at the time of updating this paper to the current version the cited references have been withdrawn \cite{Harlow:2012me} or completely rewritten \cite{Bousso:2012as}, we retain this discussion to highlight some crucial issues regarding communication and observer complementarity.} Bousso and Harlow  argue for the persistence of an information-free horizon using observer complementarity saying the various subsystems are entangled differently for different observers.\footnote{We would like to point out a confusing use of the term equivalence principle in \cite{Bousso:2012as} where the notion of observer complementarity is introduced as a response to AMPS' result in order to `save' the equivalence principle. AMPS found that there is a stress tensor at the horizon and therefore concluded that an infalling person does not fall freely. This is no more a violation of the equivalence principle than an astronaut not feeling weightless upon re-entry into the 
Earth's atmosphere is. We thank David Turton for this analogy.} Giddings had earlier argued in \cite{Giddings:2011ks,Giddings:2012bm} for non-local effects that will restore unitarity but which are ``non-violent'' such that all infalling observers have a free infall through the horizon.

Our purpose in this letter is to explicate AMPS' analysis in the language of decoherence and ask what the ÔfateÕ of an infalling wave packet is. While the main result of AMPS, that there must be information at the horizon in order to preserve unitarity, agrees with the essence of the fuzzball proposal, when phrasing the infall question in the language of decoherence we disagree with AMPS' interpretation of this structure as a firewall at which infalling observers {\em universally} burn. We first comment on the realization of the degrees of freedom as fuzzballs which are singularity-free and horizonless configurations with some very complicated structure in a region around the would-be horizon. We advocate, by the application of Occam's razor, that fuzzballs are the most conservative resolution of the information loss paradox. We then address the infall question in detail using the decoherence language. We find that depending on the width of the infalling wave packet its interaction with early and late radiation is different. While it seems unlikely that for narrow wave packets (\mbox{$E \gg T_H$} asymptotically) the interaction with early and late radiation is consistent with the picture AMPS advocate\footnote{AMPS' argument of burning is based on an infalling observer's measurement of early radiation projecting the entangled late radiation in the number operator basis as pointed out by Nomura et al. \cite{Nomura:2012sw}. For narrow wave packets such a  `measurement', which is governed by local, unitary evolution, will, if it can be performed, be very fine tuned.}, wide wave packets (\mbox{$E \sim T_H$} asymptotically) perceive a `thermal bath' of quanta which one may call a firewall. Note, however, that the interaction of low energy quanta with this `thermal bath' is not constrained to (a microscopic distance from) the horizon and is thus qualitatively different from AMPS firewall.  In the context of fuzzballs this can be seen as a macroscopic probe being able to see only the coarse grained description while a microscopic probe can perceive the structure of the fuzzballs.

\section{The Gedankenexperiment of AMPS} \label{AMPS}

AMPS start with three postulates as put forth in \cite{Susskind:1993if} based on the assumption that black hole evolution is consistent with quantum mechanics:
\begin{itemize}
\item Postulate 1 (BHC): The process of formation and evaporation of a black hole, as viewed by a distant observer, can be described entirely within the context of standard quantum theory. In particular, there exists a unitary S-matrix which describes the evolution from infalling matter to outgoing {\em Hawking-like} radiation.	
\item Postulate 2 (BHC): Outside the stretched horizon of a massive black hole, physics can be described to a good approximation by a set of semi-classical field equations.
\item Postulate 3 (BHC): To a distant observer, a black hole appears to be a quantum system with discrete energy levels. The dimension of the subspace of states describing a black hole of mass M is the exponential of the Bekenstein entropy S(M).
\end{itemize}
and add one more Postulate to this set which states:
\begin{itemize}
\item Postulate 4 (AMPS): A freely falling observer experiences nothing out of the ordinary when crossing the horizon\footnote{It is worth noting that the word ``experiences'' might leave room for different interpretations. In addition the observer-centric language can cause confusion, like the Schrodinger cat paradox has already taught us. Below we review the main points of AMPS' argument and in Section \ref{ours} we reformulate their argument in terms of local interactions of wave packets.}.
\end{itemize}

They go on to argue citing  results of \cite{Page:1993df}  that if a body in a pure state is radiating unitarily, the entanglement entropy in the radiation initially rises but at some point has to start decreasing and eventually reaches zero when the body has radiated away
completely. Moreover, there is an upper bound, known as the Page time,  when the entropy has to start decreasing: namely when half the entropy has been radiated away.              
Imagine a pure state collapsing into a black hole and emitting half its entropy in early Hawking radiation denoted by A. Unitary black hole evaporation now requires that any further outgoing quantum of radiation B has to be maximally entangled with A so that the entropy of the combined system of early and later Hawking radiation starts decreasing.
Let us introduce an infalling observer, called Alice, who encounters these outgoing Hawking quanta B close to the stretched horizon and later their partner quanta C behind the horizon. Since B is already maximally entangled with A it cannot be maximally entangled with C, the latter is however a necessary requirement for the BC system to be the vacuum state for Alice. AMPS  claim that this implies Alice encounters high energy quanta and `burns up', hence the name {\it firewall}.

In summary, the postulates 1), 2) and 4) - purity of the Hawking radiation, semi-classical behavior outside the horizon and absence of infalling drama, are mutually inconsistent and AMPS decide to give up the last one. That an information-free horizon cannot lead to a unitarity evolution of black hole evaporation has already been shown by Mathur \cite{Mathur:2009hf,Mathur:2011uj}. 
But while he proposes \cite{Mathur:2012dx,Mathur:2012zp} that the interaction of Alice with the fuzzball\footnote{In the fuzzball proposal the horizon disappears and, instead of the `state at the horizon', an infalling observer interacts with the fuzzball.} involves strong dynamics that may give rise to 
a complementarity picture that leads to free infall for high energy observers
\footnote{We will have more to comment on Mathur's \emph{fuzzball complementarity} proposal in section \ref{Fuzzballs}.}, AMPS  propose that every infalling observer burns up at the horizon.

\section{The rephrased Gedankenexperiment} \label{ours}

We wish to rephrase the `observer-centric' language of the previous section in favor of  decoherence \cite{Schlosshauer:2003zy}. Therefore, we replace the `observers' Alice and Bob by wave packets. While we will continue to use the names Alice and Bob, they should not be understood as some sentient beings but as wave packets with the usual properties of large density of states etc. to make them classical enough\footnote{Since we view Alice and Bob as wave packets we will use the word {\em it} to refer to them. We apologize to them for this rudeness.}. In this picture wave packets interact when they overlap via a local, unitary evolution and get entangled. 

\subsection{Black hole complementarity}

\begin{figure}[htbp] 
   \centering
   \includegraphics[width=0.25\textwidth]{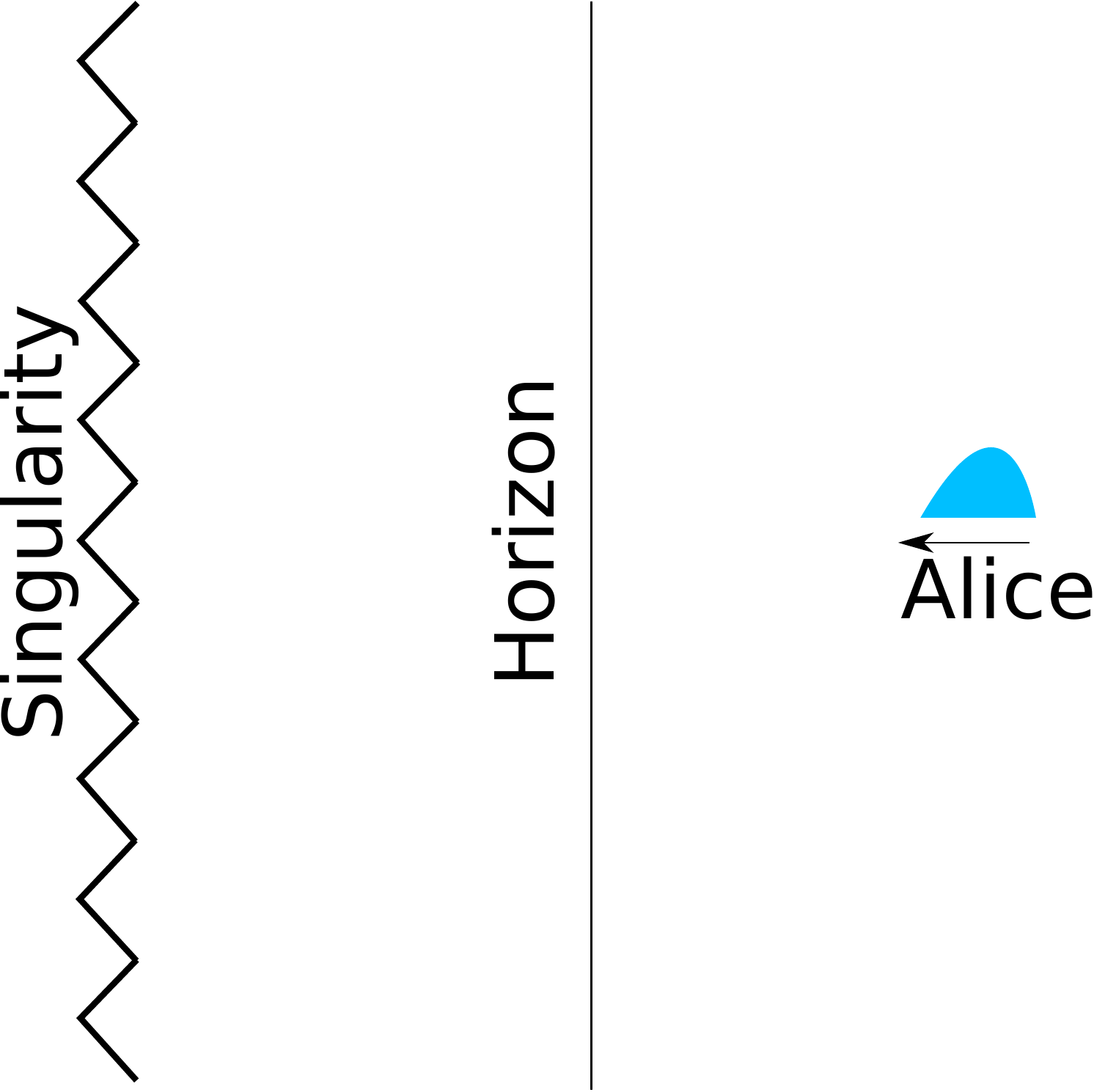} 
   \caption{A wave packet far away from the horizon evolves semi-classically.}
   \label{fig:BHCIdeaAliceAway}
\end{figure}

\begin{figure}[h!]
\begin{center}
\subfigure[]{
 \includegraphics[width=0.15\textwidth]{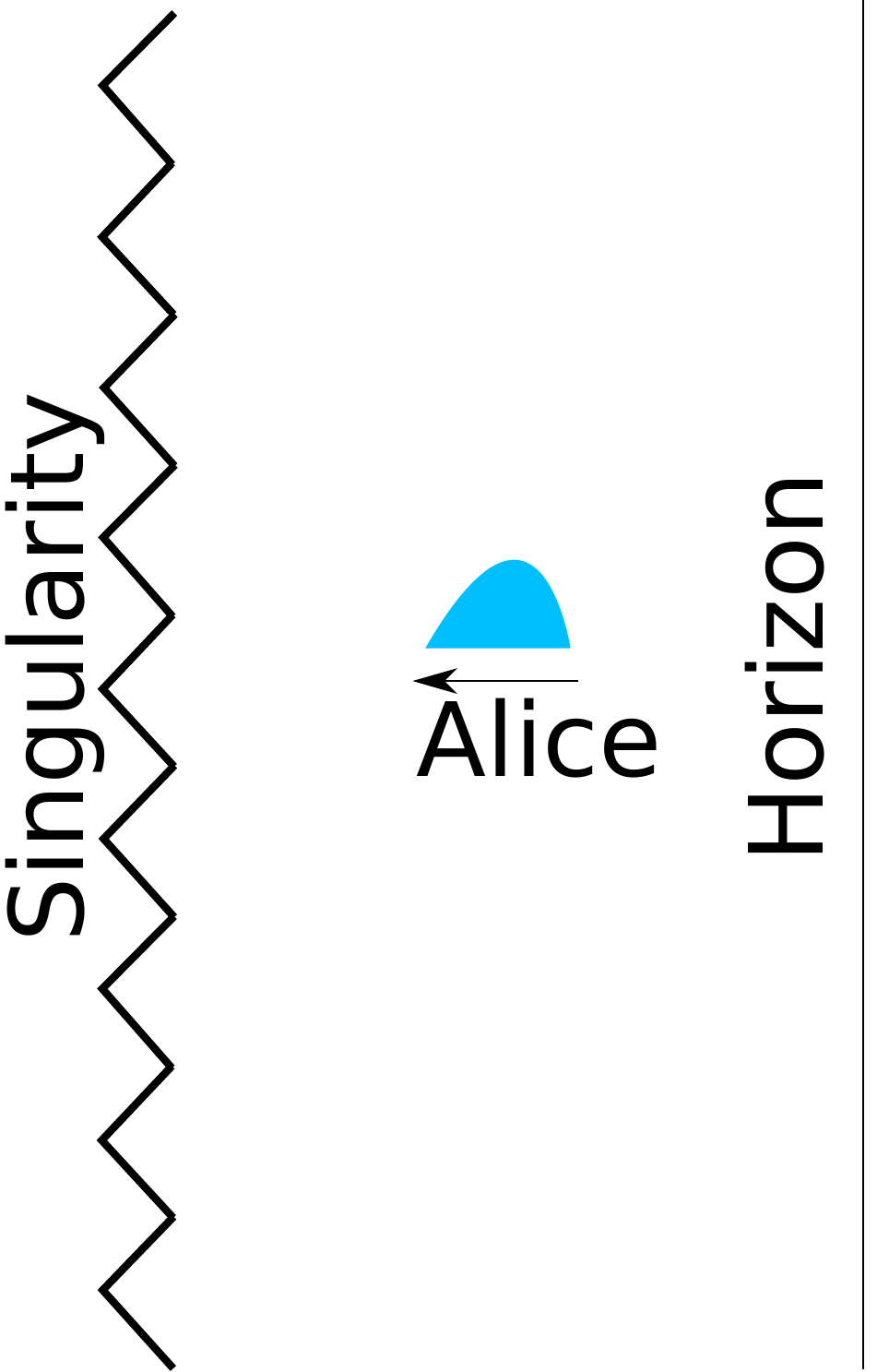}} 
\hspace{3cm}
\subfigure[]{
 \includegraphics[width=0.15\textwidth]{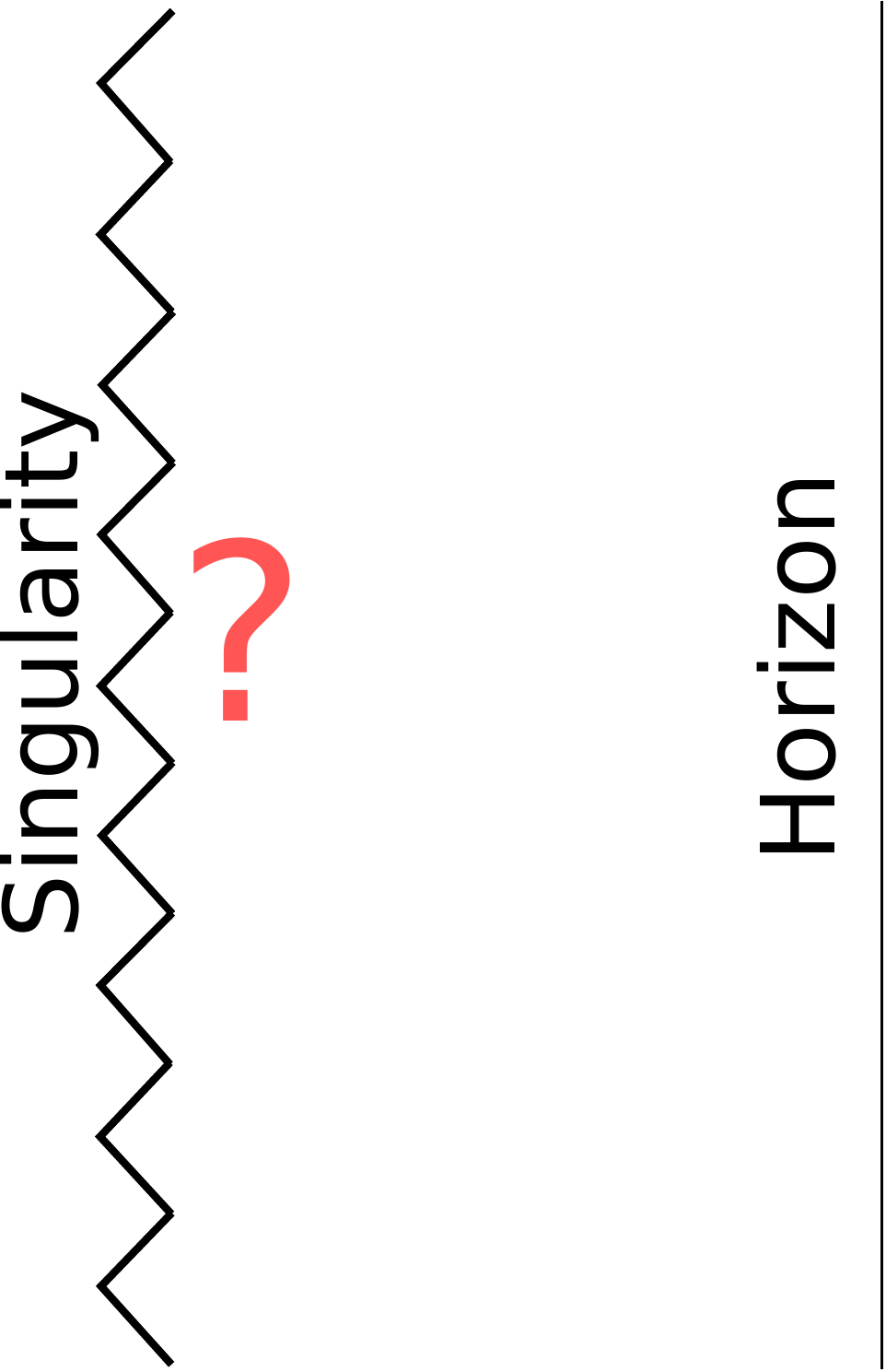}}
\end{center}
 \caption{One of the complementary descriptions is the wave packet passes through the horizon and hits a singularity.}
   \label{fig:BHCIdeaAliceClose}
\end{figure}

\begin{figure}[h!]
\begin{center}
\subfigure[]{
 \includegraphics[width=0.05\textwidth]{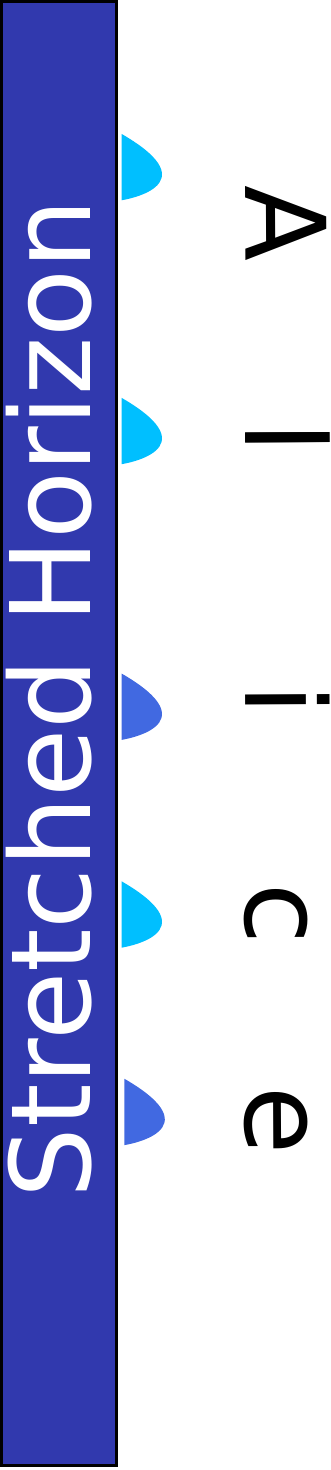}} 
\hspace{4cm}
\subfigure[]{
 \includegraphics[width=0.11\textwidth]{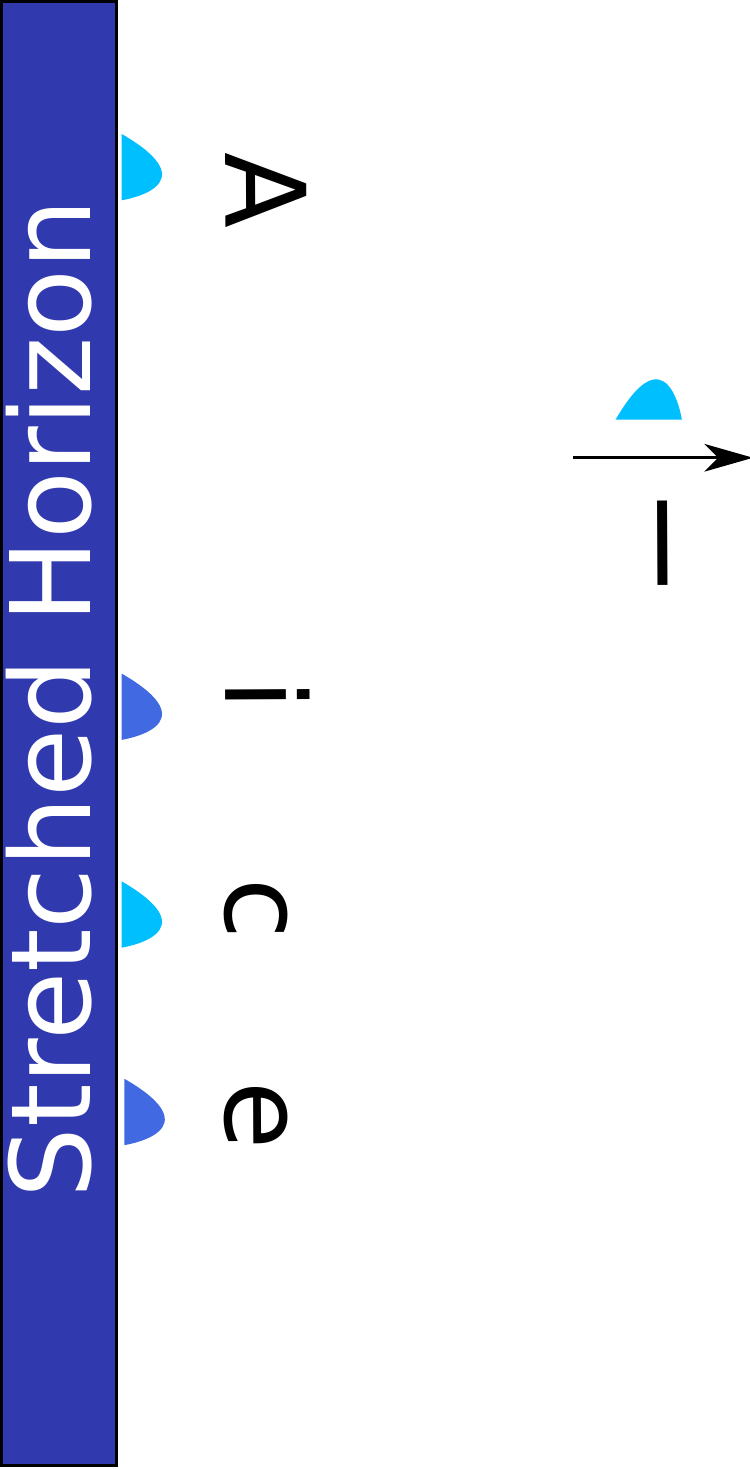}}
\end{center}
 \caption{The other complementary description is that the wave packet gets mapped onto the degrees of freedom on a membrane. This wave packet is now scrambled, looses any semblance of itself, but the information leaks out of the membrane unitarily.}
   \label{fig:BHCIdeaAliceScrambled}
\end{figure}

We want to understand the fate of a wave packet that is moving towards the black hole horizon in the black hole complementarity picture. Far from the black hole Alice is described by semi-classical evolution (figure \ref{fig:BHCIdeaAliceAway}).
When it gets close to the horizon there are two complementary descriptions: one where it passes through and then hits the singularity (figure \ref{fig:BHCIdeaAliceClose}) and another where it hits a `membrane', scrambles, and with its information being finally re-emitted unitarily, escapes to infinity (figure  \ref{fig:BHCIdeaAliceScrambled}).


While discussing in a non-observer centric language we realize that the crucial feature in black hole complementarity is that when the wave packet reaches the stretched horizon it evolves in two distinct ways. In {\em some sense}, its state gets mapped onto  {\it two copies} in separate Hilbert spaces which then evolve with different Hamiltonians\footnote{For discussion why this is consistent with no quantum cloning see \cite{Susskind:1993mu,Hayden:2007cs} and our comments below.}.

This remarkable proposal was forwarded in \cite{Susskind:1993if,Susskind:1993mu} to reconcile a pair of otherwise incompatible statements, that there is nothing from which a wave packet can {\em bounce off} at the horizon and yet {\em somehow} information must be recovered. 

In normal situations such an ad hoc prescription is not allowed as it leads to many inconsistencies and - in trying to fix them - unnecessary postulates. Copying of quantum states e.g. leads to {\it cloning}. While this does not evoke problems if the copies cannot interact as is in the case of black hole complementarity, one might still ask what it means to have two copies of a state?

This prescription is consistent if the complementary pictures are {\it dual} descriptions. For example, the state of a closed string heading towards a stack of D-branes gets mapped onto them as open string states in one picture, while in the dual description the close string continues to move on into an AdS space\footnote{We thank Samir Mathur for pointing this out to us.}. For such dual descriptions there is no issue with cloning. However, the Hamiltonian evolutions of the states need to be consistent since, at the end of the day, in one description, the closed string emerges out of the AdS as a closed string in flat space and in the other description the open strings leave the D-brane as a closed string. They must be in the {\it same state}.

However, black hole complementarity is not a duality. This conclusion can be drawn from the completely different outcomes in the two complementary pictures.\footnote{We note in passing that AdS/CFT does not provide a solution to the information paradox: this duality requires the same evolution of two copies of information but in the presence of a black hole in the infra-red the evolution of these two copies is different and there is no duality.} If it is not a duality then one can ask {\it what} it is and what is this operation that makes two copies of the states\footnote{If one is uncomfortable with the language of making two copies of the states one can say it is two descriptions of the same state which evolve differently.}? 
We will comment on a recent proposal of Mathur where an \emph{energy scale dependent} complementarity can arise in the context of fuzzballs in Section \ref{Fuzzballs}. For now we continue with our non observer-centric description.

\subsection{The firewall argument}\label{AMPSwave}

\begin{figure}[h!]
\begin{center}
\subfigure[]{
 \includegraphics[width=0.4\textwidth]{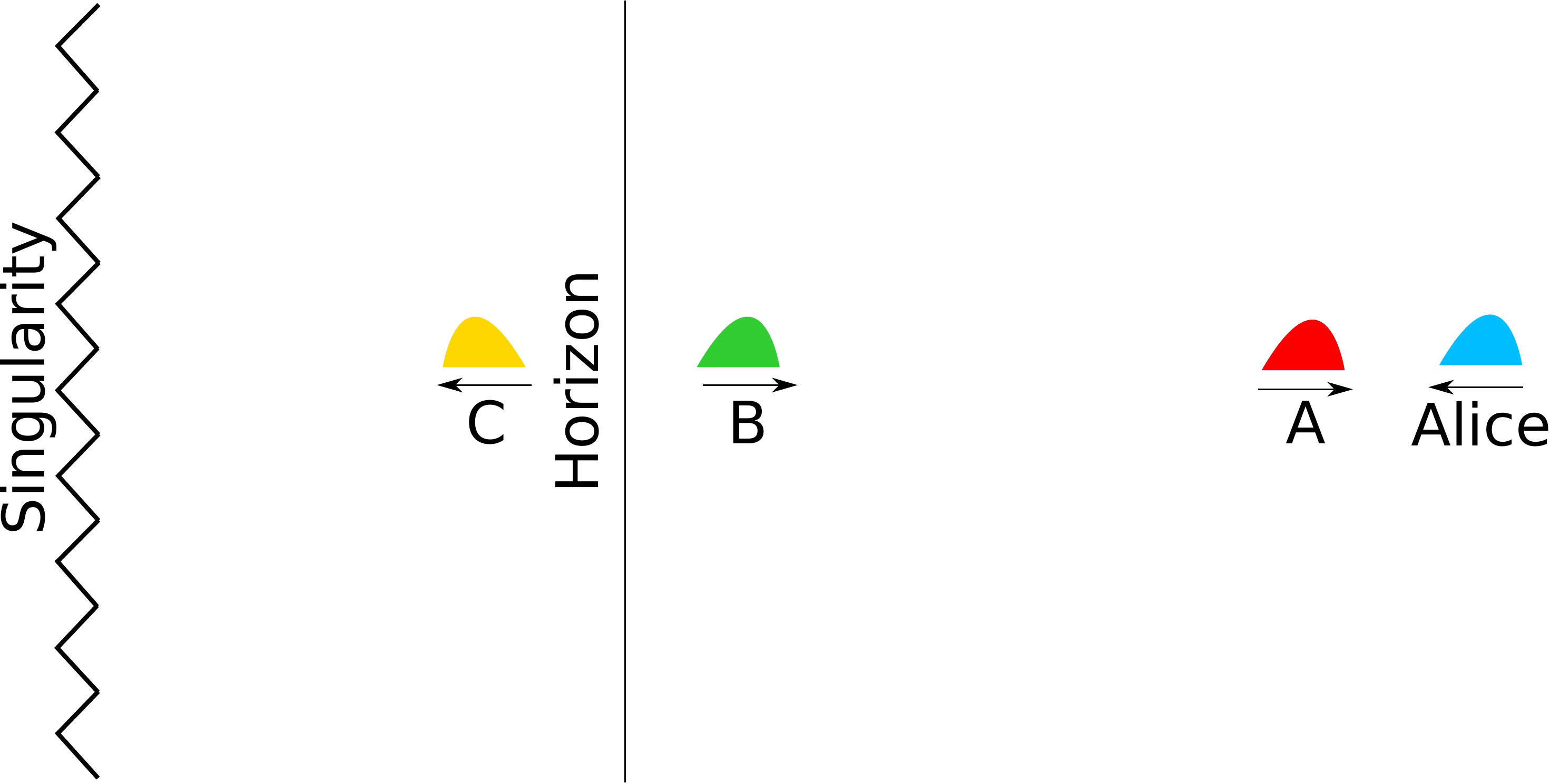}} 
\hspace{0.8cm}
\subfigure[]{
 \includegraphics[width=0.4\textwidth]{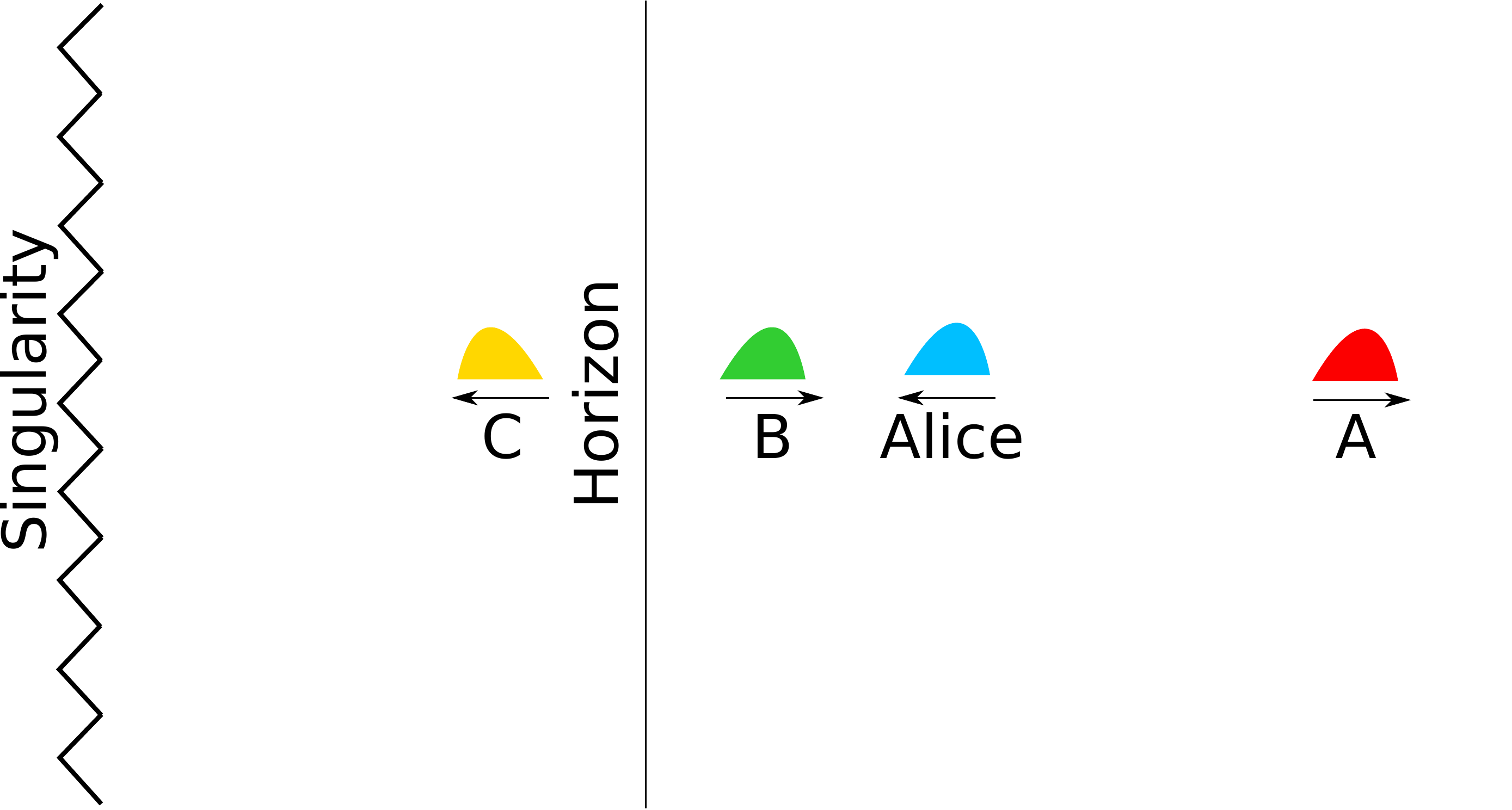}}
\hspace{0.8cm}
\subfigure[]{
 \includegraphics[width=0.4\textwidth]{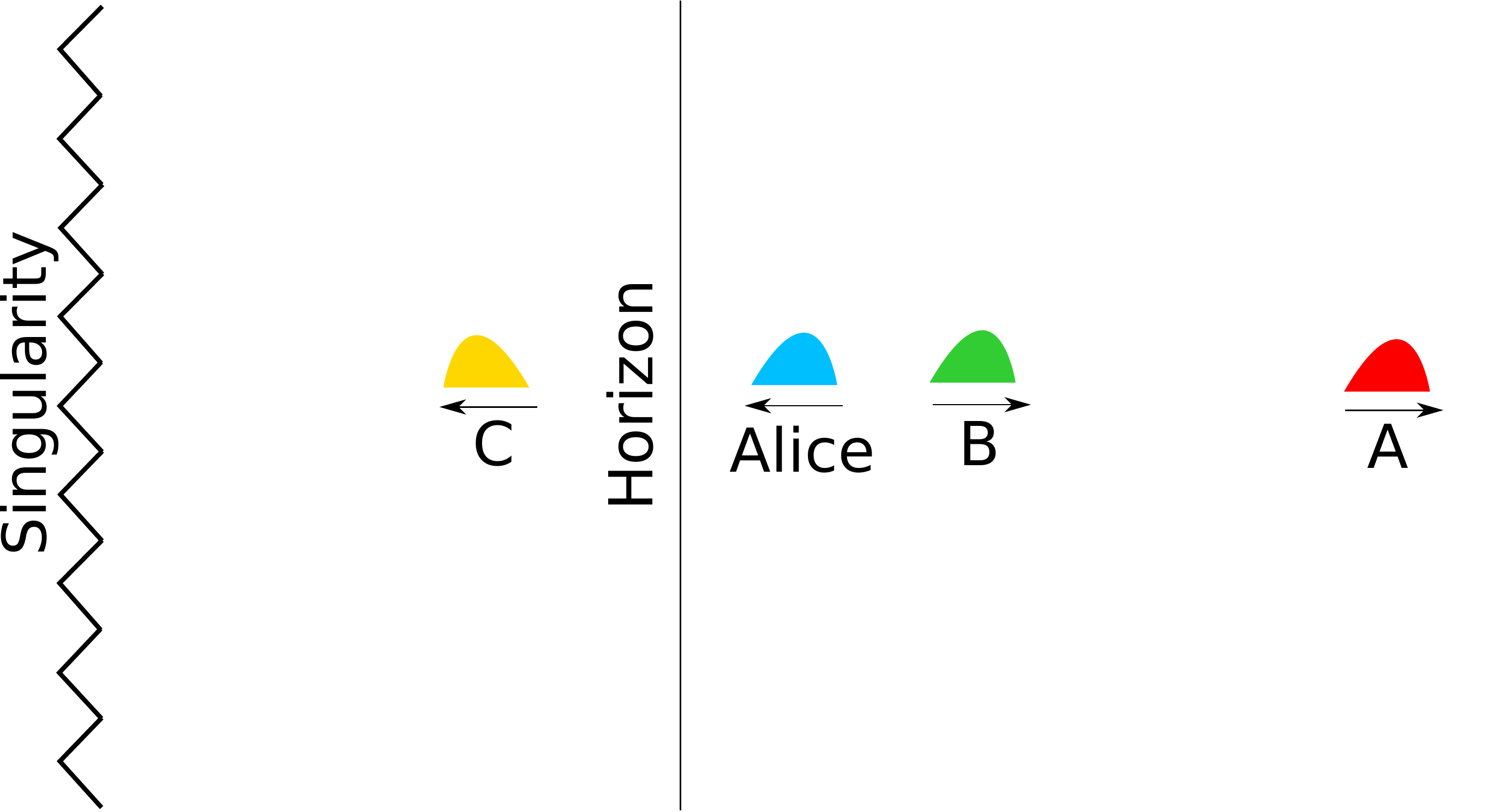}}
\end{center}
 \caption{Away from the horizon Postulate 2 tells us that the usual rules of quantum mechanics work. When wave packet `Alice' crosses the wave packets `A' and `B' they will get entangled successively.}
   \label{fig:BHCOne}
\end{figure}

Now let us look at the AMPS Gedankenexperiment in terms of these wave packets. We begin with the unitary evolution process of Alice away from the black hole in the causal future of early radiation A. While for clarity we have drawn only one wave packet A in figures \ref{fig:BHCOne}, it should be thought of as many wave packets: when Alice interacts with A, it interacts in fact with many such wave packets successively. How strong the interaction is and how much entanglement will be generated between Alice and A in the process is governed by local unitary dynamics and the properties of the wave packet Alice, e.g. what frequencies is the wave packet supported on. This will cause the crucial difference in Alice's `experience' which we alluded to in Section \ref{AMPS}. We will come back to this issue in  Section \ref{Burn}. Regardless of this, in each encounter there is some mixing between Alice and wave packets in A.

When Alice gets closer to the horizon it interacts with wave packet B. Since B is just a blue shifted version of A and Alice is also blue shifted, this interaction is stronger than any of the Alice-A interactions. 

\begin{figure}[h!]
\begin{center}
\subfigure[]{
 \includegraphics[width=0.4\textwidth]{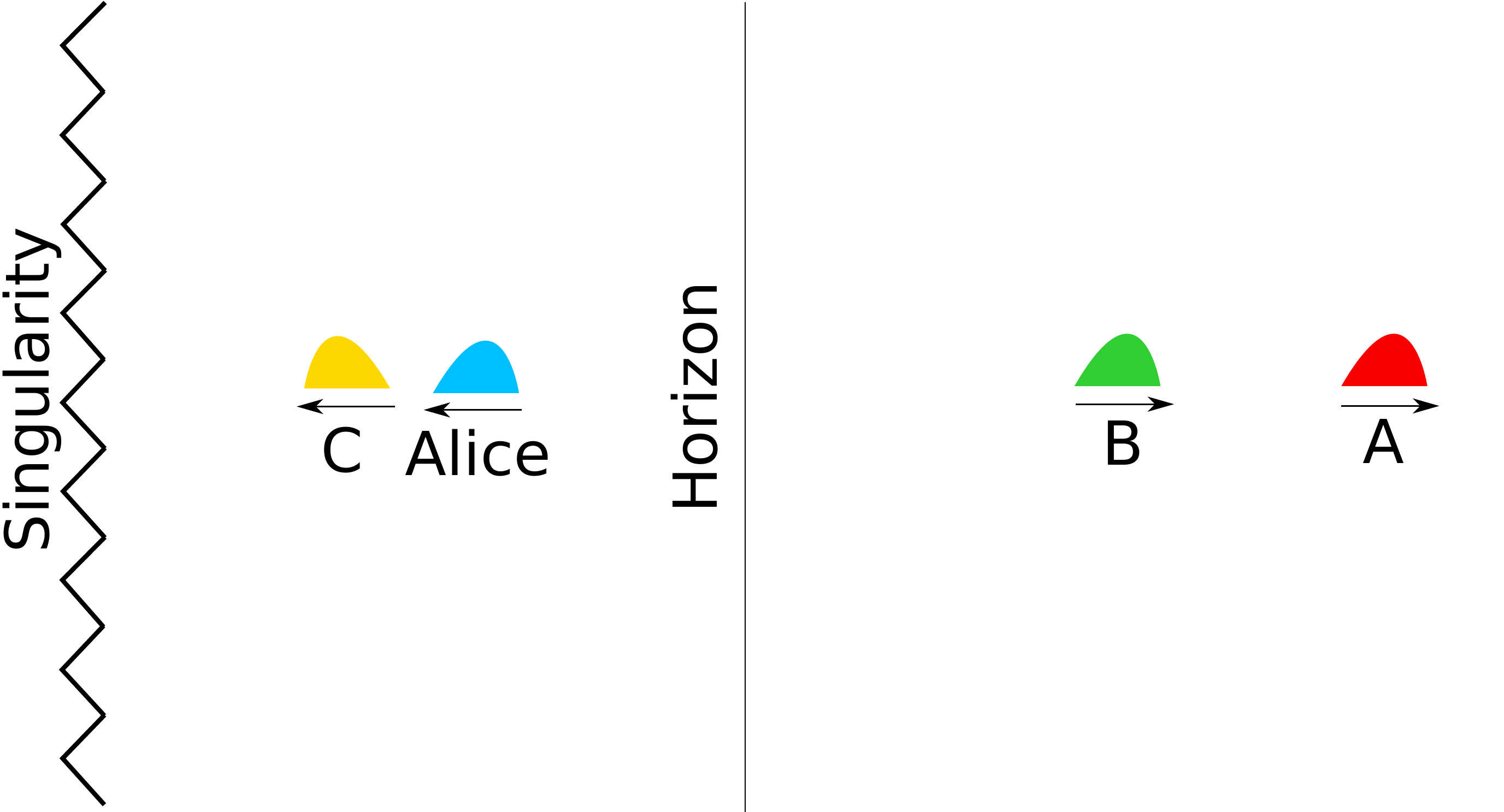}} 
\hspace{1.8cm}
\subfigure[]{
 \includegraphics[width=0.4\textwidth]{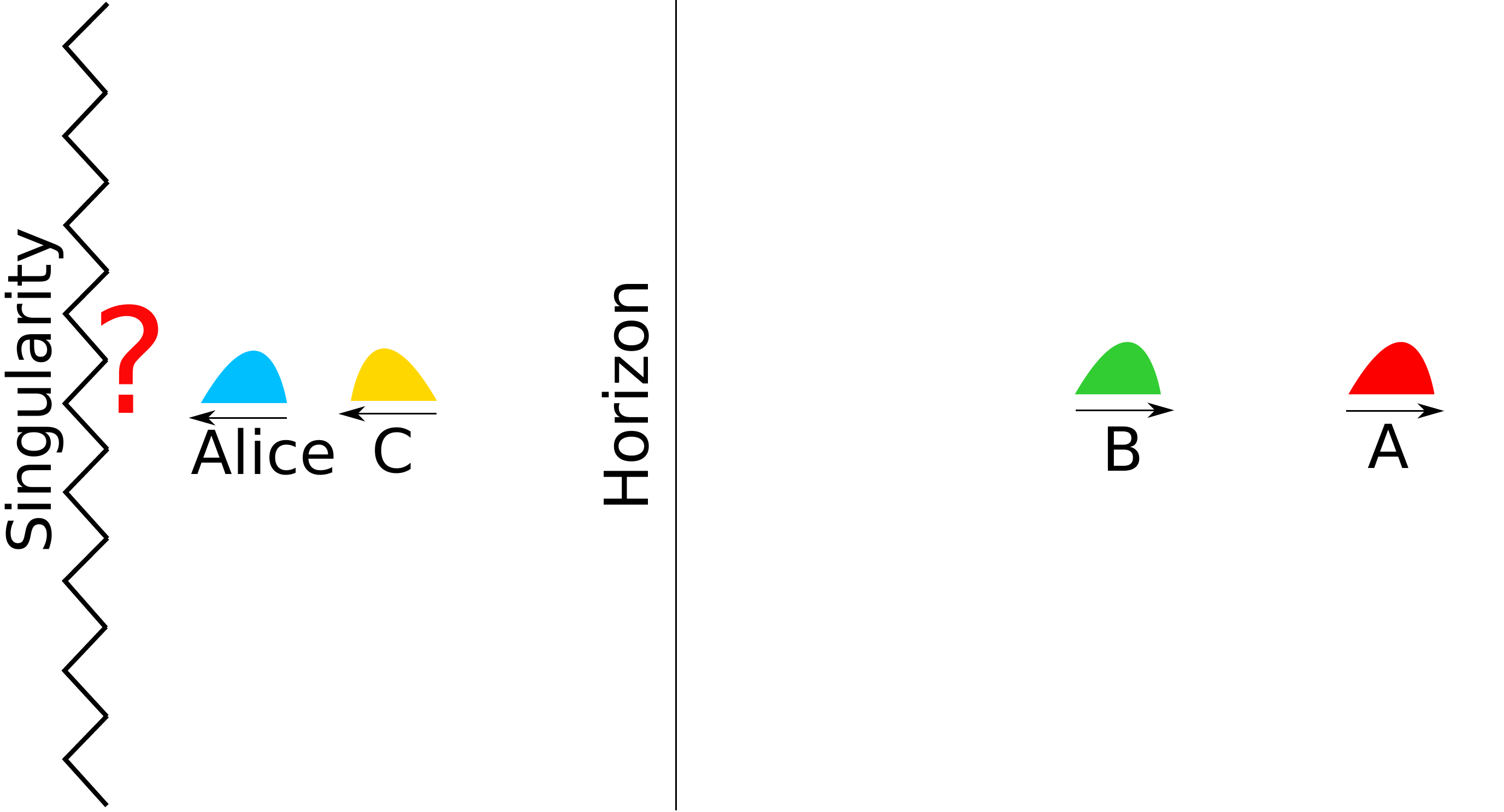}}
\end{center}
 \caption{ In one complementary picture Alice having interacted with A and B falls through the horizon. Its interaction with C will depend on whether B was entangled with A or C. It then falls into the singularity.}
   \label{fig:BHCPassing}
\end{figure}

\begin{figure}[h!]
\begin{center}
\subfigure[]{
 \includegraphics[width=0.35\textwidth]{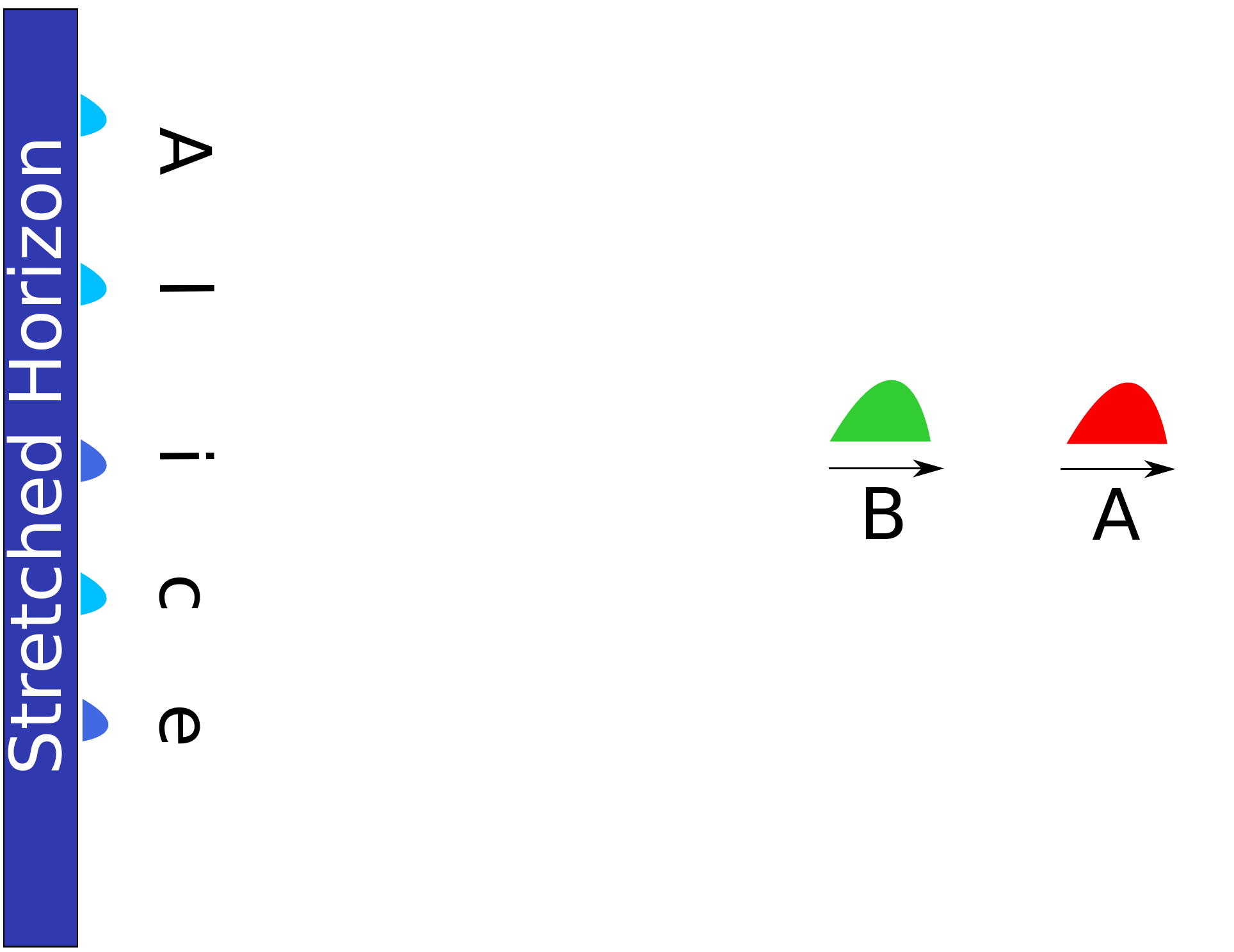}} 
\hspace{2cm}
\subfigure[]{
 \includegraphics[width=0.35\textwidth]{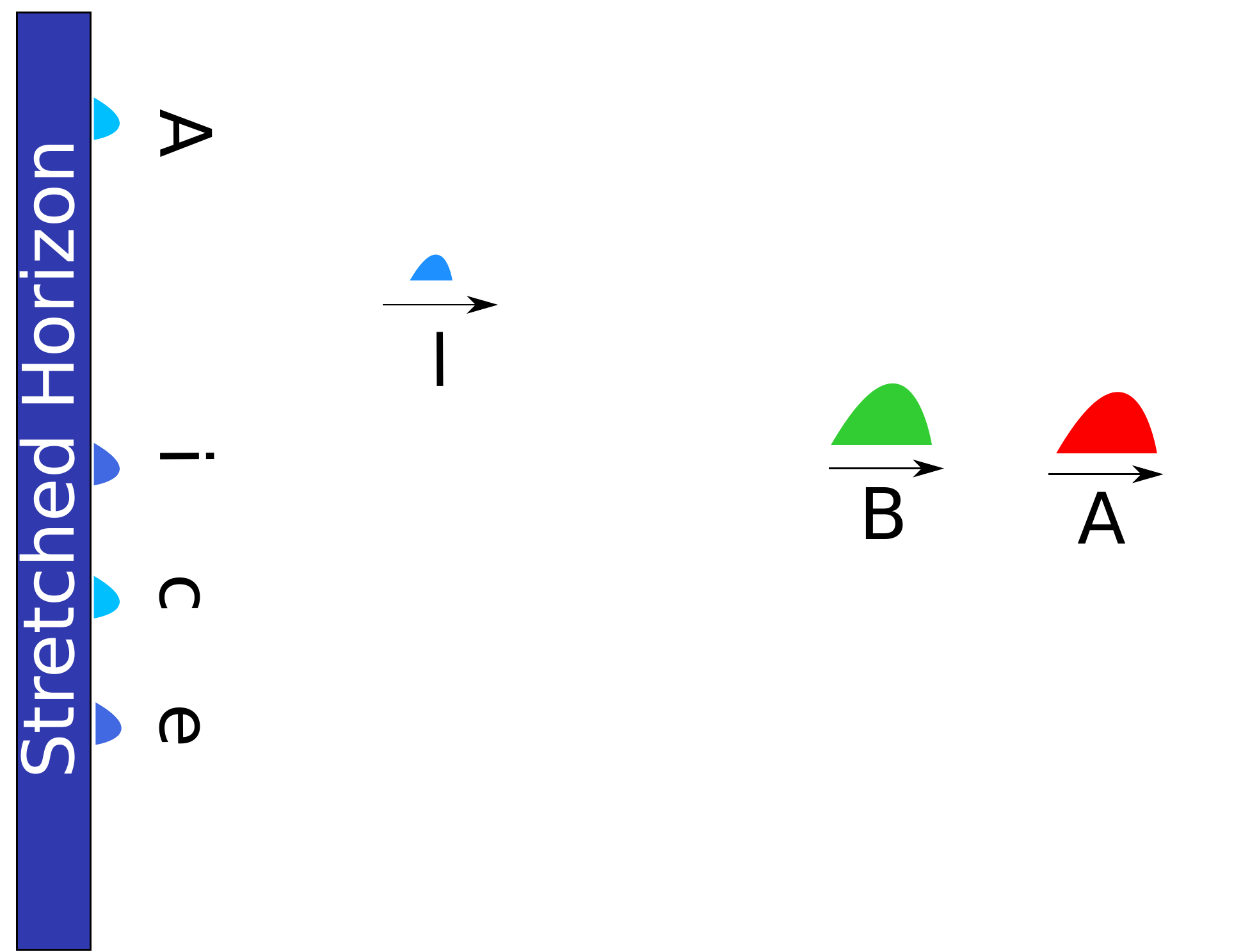}}
\end{center}
 \caption{In the other complementary picture Alice's state after having interacted with AB is registered on the membrane. It gets scrambled but eventually leaks out unitarily.}
   \label{fig:BHCHitting}
\end{figure}

After these interactions Alice heads towards the stretched horizon. At this point, it is worth observing that Alice's interactions with A and B are encoded in Alice and in the AB system. We have not talked about any measurements or observations. We will have something to say about this in  Section \ref{Burn}.
When Alice reaches the (stretched) horizon we make use of black hole complementarity: 
According to one picture Alice (entangled with AB)  goes through the horizon and encounters wave packet C, the infalling Hawking pair of B, with which it interacts and mixes and eventually hits the singularity (figure \ref{fig:BHCPassing}).\footnote{While one may argue that the singularity will get resolved at Planck scale it is not clear how this picture can be modified to make black hole complementarity an actual duality.} In the complementary picture Alice's (entangled with AB) state gets mapped onto the thermal membrane at the stretched horizon where it gets scrambled  
and eventually re-emitted unitarily as radiation to infinity (figure \ref{fig:BHCHitting}).

Unitarity of black hole evaporation requires B to be maximally entangled with A while free infall requires B to be maximally entangled with C. A system cannot be maximally entangled with two distinct systems. Alice interacting with system AB will have different evolutions depending on whether B was maximally entangled with A or C to begin with.  The state of this wave packet will then be recorded on the stretched horizon in one of the black hole complementarity pictures. In the other complementary picture it falls through the horizon, and interacts with C. The interaction of Alice with C will have a different evolution depending on whether B was maximally entangled with A or C to begin with. Eventually in this complementary picture Alice hits the singularity.

AMPS argue that since B is maximally entangled with A, Alice's encounter with B is fatal for it (which in our language should be read as its wave packet changing so much that it does not resemble its former self). Thus they advocate that black hole complementarity has to be given up. However, while it is clear that Alice's state will be different from what one would have thought without the benefit of AMPS Gedankenexperiment, it is not immediately clear by how much. The answer to this will depend on the properties of the wave packet Alice and the Hamiltonian. 
We thus see the possibility of a \emph{scale dependent} complementarity that depends on the properties of the infalling wave-packet. 
Such an idea was recently proposed by Mathur and we will talk about this more in Section~\ref{Fuzzballs} and Section~\ref{Burn}.

\subsection{Observer complementarity: Alice in Wonderland} \label{responses}

In follow-up papers to the AMPS argument Bousso \cite{Bousso:2012as} and Harlow \cite{Harlow:2012me} have argued, using  {\em observer complementarity}, that the interpretation of strong sub-additivity by AMPS is incorrect. They claim that different observers can find different answers to the question of whether B is maximally entangled with A or C without leading to contradictions. 

We list the postulates stated explicitly in \cite{Harlow:2012me}:
\begin{itemize}
\item Postulate 1 (H): From the point of view of observers asymptotically far away from the black hole,
its formation and evaporation may be described by a unitary S-matrix. In the intermediate state where the black hole exists, its microscopic entropy is given by the Bekenstein-Hawking formula $S = \frac{A}{4G}$.

\item Postulate 2 (H): No observer sees violations of low energy effective field theory away from a
stretched horizon.

\item Postulate 3 (H): If two observers can causally communicate the results of their experiments, they must agree on the results of those experiments.

\item Postulate 4 (H): An observer freely falling through a sufficiently large black hole horizon from
some finite distance away in Schwarzschild units will experience no ill effects in
doing so.

\end{itemize}
Furthermore Postulate 3 of section \ref{AMPS} is implicitly assumed\footnote{We thank Daniel Harlow for an explanation of observer complementarity.}.
  
In \cite{Bousso:2012as} it is argued that since different observers have different causal diamonds they have their `own theory'. 
Although an outside and an infalling observer, both, need to find the early Hawking radiation A to be consistent with unitary evolution, the infalling observer  `relinquishes' the possibility of measuring the full S-matrix and is therefore `free to claim' that A is not entangled with B. After crossing the horizon the infalling observer then concludes that B and C are entangled and is safe. 

While \cite{Bousso:2012as} points out that the possibility of communication between the two observers could lead to a contradiction in the outside diamond, it is argued that it should not be possible for an infalling observer to send a message and report the absence of a drama at the horizon to the outside observer. 
Ref.~\cite{Harlow:2012me} follows through with a calculation arguing that the possibility of a communication between the infalling and the outside observer is not realized. An infalling observer who has an upper bound on what temperatures are `unhealthy' i.e. will burn him/her, cannot process the information of the encounter with such an unhealthy quantum B and send it to an asymptotic observer before getting to the stretched horizon.
 
There is a potential problem with the above argument. If infalling Alice is to semi-classically communicate  her experiences to Bob then they have to have a predetermined algorithm. For example they may decide that Alice in her frame sends a red signal if she encounters B entangled with A and a yellow signal if she encounters B entangled with C. 
When Bob receives her signal he will use his effective field theory transformation rules to reverse the distortion that Alice's signal experienced while traveling to him. If this transformation yields a red signal Bob may conclude that Alice encountered B entangled with A. However, if the physics close to the horizon is so different that Alice may encounter B entangled with C the yellow signal she sends may, while traveling to Bob, become distorted differently such that Bob's transformation rules may tell him that the signal Alice send was red. Thus, allowing  physics close to the horizon to be so different for different observers would  lead to a communication breakdown anyways and thus also to information loss even before the horizon is reached. 

There are some other problems with using observer complementarity to argue for an information-free horizon also because of which, by the time of updating to the current version of this article, \cite{Harlow:2012me} has been withdrawn and \cite{Bousso:2012as} has been modified substantially.

\section{Fuzzballs} \label{Fuzzballs}

Black hole complementarity tried to reconcile the two opposing ideas of unitarity and free infall at the horizon.  History has shown us that attempts to reconcile contradicting fundamental notions on the black hole solution cannot be forced upon without leading ad absurdum. Mathur \cite{Mathur:2009hf,Mathur:2011uj} has recently shown that to be able to describe black hole evaporation by a unitary S-matrix for an asymptotic observer the traditional picture has to give way to one where the state at the horizon is not the vacuum state for an infalling observer. AMPS have also agreed with this result.

This picture was long ago proposed by Mathur \cite{Lunin:2002qf,Mathur:2005zp} and incorporated in the fuzzball community. 
According to the fuzzball conjecture the true microstates of quantum gravity are singularity-free and horizonless solutions. While one would expect that most of the typical fuzzballs require a fully stringy description in the core region and are therefore not describable in terms of supergravity, some of them may actually admit a supergravity description in terms of smooth geometries\footnote{Even if this is not the case for all fuzzballs, those are the ones that are possible to construct with current technology and they are very useful to probe the properties of black holes.}.
In either case, since there is no horizon and no singularity there is no information loss. The radiation is emitted from the surface of the fuzzball which need not be, and most likely is not, constrained to within a Planck length from the horizon. Indeed, a family of non-typical near-extremal black hole microstates ~\cite{Bena:2012zi}, recently constructed within supergravity, indicates that the surface of the fuzzball fluctuates from one solution to the other.
For  reviews on fuzzballs we  refer the reader to \cite{Bena:2004de,Mathur:2005zp,Skenderis:2008qn,Mathur:2008nj,Balasubramanian:2008da,Chowdhury:2010ct}
and to see how radiation from fuzzballs carries information see \cite{Chowdhury:2007jx,Avery:2010hs}.

It is then interesting to ask what happens to an observer falling into a fuzzball.
Mathur has recently addressed this question proposing \emph{fuzzball complementary} \cite{Mathur:2012dx,Mathur:2012zp} based on \cite{Czech:2012bh,Czech:2012be} which can be viewed as an approximate form of complementarity distinguishing between different energy scales of infalling observers. According to this proposal when a high energy wave packet ($E \gg T_H$) hits a typical fuzzball, it excites the collective modes of the latter. 
While the details of this process are not describable with the current technology, it is possible to make a coarse-grained approximation where infall into a typical fuzzball can be replaced by infall into a black hole. 
The wave packet hitting the singularity should be interpreted as thermalization of the collective modes.
Mathur refers to this complementary picture for fuzzballs as \emph{approximate} because unlike black hole complementarity the picture of free infall  only works for $E \gg T_H$.

A realization of  fuzzballs, inherently stringy objects, giving a complimentary description of black holes would be a fulfillment of the prediction made in \cite{Susskind:1993mu}: {\em It is our view that black hole complementarity is not derivable from a conventional local quantum field theory. It seems more likely that it requires a radically different kinematical description of physics at very high energy, such as string theory. }

\section{Is Alice burning or fuzzing?} \label{Burn}

The results of \cite{Almheiri:2012rt,Susskind:2012rm,Susskind:2012uw} support the basic idea of the fuzzball proposal that there has to be an order one correction (i.e. not suppressed in $M^{-1}$) to the black hole horizon to preserve unitarity. However, they claim that infalling observers of all energy scales burn up at the horizon as opposed to the idea of fuzzball complementarity mentioned in Section \ref{Fuzzballs}. In this Section we will give some evidence in support of the fuzzball complementarity picture.

Nomura et al claim in \cite{Nomura:2012sw} that AMPS' conclusion that an infalling observer sees a firewall is incorrect based on the following very interesting reasoning. AMPS' claim 
that  since the state of the final Hawking radiation is pure it can be written as an early and late part
\be
\ket{\Psi} = \sum_i \ket{\psi_i}_E \otimes \ket{i}_L\,,
\ee
where $\ket{i}_L$ is an arbitrary complete basis for late radiation and $\ket{\psi_i}_E$ is a state in the early Hawking radiation. After Page time the Hilbert space of the early radiation will be much larger than that of the late radiation and so, for typical states $\ket{\Psi}$, the reduced density matrix describing the late-time radiation is close to the identity. Thus one can construct operators acting on early radiation whose action on $\ket{\Psi}$ is
\be
P_i \ket{\Psi} \propto \ket{\psi_i}_E \otimes \ket{i}_L\,.
\ee
While this is true, AMPS claim that Alice can make measurements on early radiation tantamount to projection into the eigenvector of the number operator. In \cite{Nomura:2012sw} an objection is raised to  this stating that `the existence of the projection operator for an arbitrary $i$  does not imply that a measurement in a sense that it leads to a classical world can occur to pick up the corresponding state'. In other words measurement is a dynamical process dictated by unitary evolution of the state. 

Translated in our wave packet language this means that when Alice passes through early radiation A it cannot {\em choose} which basis it projects onto. This is because the interaction of Alice and A is governed by a local Hamiltonian. Let the typical energy of quanta in A be $T_H$ which is of the order of the Hawking temperature. Then we have two different kind of scenarios:
\begin{itemize}
\item Alice is a wave packet with support on energies $E \gg T_H$ and is thus smaller than typical wave packets in A. In this case Alice interaction with A does not project onto the number operator basis in any typical interaction and it cannot  `predict' the number of quanta in a mode of B.
\item Alice is a wave packet with support on energies $E \sim T_H$ and thus is of the same size as wave packets in A. In this case a typical interaction of them will project A onto the number operator basis.  When Alice falls in, it will encounter B and will be able to `predict' the number of quanta in a mode of B.
\end{itemize}
The remaining case of $E \ll T_H$ is not so interesting as for such wave packets the wavelength is bigger than the black hole. Such wave packets are reflected off the potential barrier surrounding the black hole. 

We see that we clearly have two different scenarios but what is also very interesting is that we also have two different expectations on what happens to infalling wave packets.

\begin{itemize}
\item Mathur claims that for quanta of energy $E \gg T_H$ there is approximate complementarity (approximate in that it does not apply to wave packets of energy $E \sim T_H$) as explained in the previous section. Such wave packets can have free infall. 
\item AMPS claim that an infalling observer encounters high energy quanta in the number basis near the horizon. While this seems  not to be true for $E \gg T_H$\footnote{It is not immediately clear that a local wave packet Alice cannot sample all of A on the sphere in some coherent way and then fall in and encounter B projected on a number basis. However typical interactions will not have this effect.} it is true for $E \sim T_H$. Thus, such a wavepacket will not have free infall. 
\end{itemize}

Based on these observations we have the following interesting picture of fuzzballs. High energy wave packets pass through the Hawking-like quanta without being affected much and hit the bottom of the fuzzball.
Based on \cite{Czech:2012be,Czech:2012bh}, these wave packets will excite the collective modes of fuzzballs. For typical fuzzballs an infalling wave packet with asymptotic energy $E \gg T_{H}$ experiences an effective geometry until it gets closer to the would-be singularity, whereas for an infalling wave packet with asymptotic energy $E \sim T_{H}$ spacetime ends before the horizon where it bounces off the Hawking quanta and the 'fuzz' (microstate structure at the scale of the horizon).\footnote{This picture is not very different from a wave packet heading towards a burning piece of coal. A wave packet  of thermal energy undergoes Brownian motion by being scattered off the outgoing radiation, while a wave packet of energy much greater than the thermal energy manages to travel on a `classical trajectory' and hit the piece of coal. However, the analogy stops here because there is no nice approximate complementarity picture for a hot piece of coal.}. While we focused on the extreme limit of $E\sim T_H$ and $E\gg T_H$ which yield dramatic to free infall, respectively, we expect intermediate wave packets to have an experience ranging between these two extremes.

So finally, in response to the question `Is Alice burning or fuzzing?' one should ask back what Alice is made of or sing  the song `Alice, Alice, Who the **** is Alice?' \cite{Alicesong}.

\section*{Acknowledgments}

This paper is based on many discussions we had with many people. We express our gratitude especially to Marco Baggio, Iosif Bena, Jan de Boer, Sheer El-Showk, Samir Mathur, Masaki Shigemori, David Turton, Dieter van den Bleeken and Bram Wouters. We would like to thank Daniel Harlow for helpful discussion and Finn Larsen for comments
on an earlier version of the paper. The work of BDC is supported by the ERC Advanced Grant 268088-EMERGRAV. AP is supported by DSM CEA-Saclay. BDC would like to thank the CEA Saclay for hospitality where most of this work was carried out.

\bibliographystyle{toine}
\bibliography{Papers.bib}

\end{document}